\documentclass[twocolumn,prb,showpacs,showkeys,superscriptaddress]{revtex4}

\usepackage{bm}
\usepackage{makeidx}
\usepackage{amssymb}
\usepackage{xspace}
\usepackage{amsmath}
\usepackage{graphicx}
\usepackage[cp1250]{inputenc}
\usepackage{enumerate}
\usepackage[ps2pdf,hyperindex=true,final=true,colorlinks=true,pagebackref=false,bookmarks=true,bookmarksopen=true,bookmarksnumbered=true]{hyperref}
\newcommand{\Cu}{\ensuremath{\mathrm{Cu}}}
\newcommand{\Ni}{\ensuremath{\mathrm{Ni}}}
\newcommand{\Nb}{\ensuremath{\mathrm{Nb}}}

\newcommand{\Al}{\ensuremath{\mathrm{Al}}}
\newcommand{\Si}{\ensuremath{\mathrm{Si}}}

\renewcommand{\O}{\ensuremath{\mathrm{O}}}

\newcommand{\ZeroPi}{{\ensuremath{0\text{-}\pi}}\xspace}

\begin{document}
\title{Magnetic interference patterns in \ZeroPi SIFS Josephson junctions: effects of asymmetry between $0$ and $\pi$ regions}
\author{M. Kemmler}\email{kemmler@pit.physik.uni-tuebingen.de}
\affiliation{%
  Physikalisches Institut -- Experimentalphysik II and Center for Collective Quantum Phenomena,
  Universität Tübingen,
  Auf der Morgenstelle 14,
  D-72076 Tübingen, Germany
}
\author{M. Weides}
\altaffiliation{Current address: Department of Physics, University of California, Santa Barbara, CA 93106, USA}
\affiliation{%
  Institute of Solid State Research and JARA--Fundamentals of Future Information Technology, Research Centre Juelich, 52425 Juelich, Germany%
}

\author{M. Weiler}
\author{M. Opel}
\author{S. T. B. Goennenwein}
\affiliation{Walther-Meissner-Institut, Bayerische Akademie der
Wissenschaften, 85748 Garching, Germany}

\author{A. S. Vasenko}
\affiliation{%
LPMMC, Université Joseph Fourier and CNRS, 25 Avenue des Martyrs, BP 166, 38042 Grenoble, France}
\author{A. A. Golubov}
\affiliation{%
Faculty of Science and Technology and MESA$^+$ Institute for
Nanotechnology, University of Twente, 7500 AE Enschede, The
Netherlands}
\author{H. Kohlstedt}
\affiliation{%
  Nanoelektronik, Technische Fakultät, Christian-Albrechts-Universität zu Kiel, D-24143 Kiel, Germany%
}
\author{D. Koelle}
\author{R. Kleiner}
\author{E. Goldobin}
\affiliation{%
  Physikalisches Institut -- Experimentalphysik II and Center for Collective Quantum Phenomena,
  Universität Tübingen,
  Auf der Morgenstelle 14,
  D-72076 Tübingen, Germany
}

\date{\today}
\begin{abstract}
We present a detailed analysis of the dependence of the critical
current $I_c$ on the magnetic field $B$ of $0$, $\pi$, and \ZeroPi
superconductor-insulator-ferromagnet-superconductor Josephson
junctions. $I_c(B)$ of the $0$ and $\pi$ junction closely follows a
Fraunhofer pattern, indicating a homogeneous critical current
density $j_c(x)$. The maximum of $I_c(B)$ is slightly shifted along
the field axis, pointing to a small remanent in-plane magnetization
of the F-layer along the field axis. $I_c(B)$ of the \ZeroPi
junction exhibits the characteristic central minimum. $I_c$
however has a finite value here, due to an asymmetry of $j_c$ in the $0$
and $\pi$ part. In addition, this $I_c(B)$ exhibits asymmetric
maxima and bumped minima. To explain these features in detail, flux penetration being different in the $0$ part and the $\pi$ part needs to be taken into account. We discuss this asymmetry in relation to the magnetic properties of the F-layer and the fabrication technique used to produce the \ZeroPi junctions.
\end{abstract}
\maketitle
\section{Introduction}\label{Sec_Intro}
While predicted more than 30 years
ago\cite{Bulaevskii77,Buzdin1982}, due to the severe technological requirements, the experimental study of $\pi$
Josephson junctions became an intense field of research only
recently.
Super\-con\-duc\-tor-fer\-ro\-mag\-net-su\-per\-con\-duc\-tor (SFS)
Josephson junctions were successfully fabricated and
studied\cite{Ryazanov01,Sellier04,Blum02,Bauer04}. SFS junctions
however typically exhibit only very small (metallic) resistances
$R$, making this type of junctions less suitable for the study of
dynamic junction properties as well as for applications, where
active Josephson junctions are required. To overcome this problem,
an additional insulating (I) layer can be used to increase $R$,
although at the expense of a highly reduced critical current density
$j_c$\cite{Kontos02, Weides06, Bannykh09, Sprungmann09}.

In a SFS or SIFS junction the proximity effect in the ferromagnetic
layer leads to a damped oscillation of the superconducting order
parameter in the F-layer. Thus, depending on the thickness $d_F$ of
the F-layer, the sign of the order parameters in
the superconducting electrodes may be equal or not. While in the
first case a conventional Josephson junction (a ``$0$ junction'')
with $I_s=I_c\sin(\mu)$ is realized, in the latter case a ``$\pi$
junction'' is formed where the Josephson current $I_s$ obeys the
relation $I_s=I_c\sin(\mu+\pi)=-I_c\sin(\mu)$. Here $I_c\geq0$ is the
junction critical current and $\mu$ is the phase difference of the
order parameters in the two electrodes.

The combination of a $0$ and a $\pi$ part within a single
Josephson junction leads to a ``\ZeroPi'' Josephson junction.
Depending on several parameters of the $0$ and $\pi$ part, a
spontaneous fractional vortex may appear at the \ZeroPi
boundary\cite{Bulaevskii1978}. In case of long junctions with length
$L\gg\lambda_J$ the vortex contains a flux equal to a half of a flux
quantum $\Phi_0\approx2.07 \times 10^{-15}\,\rm{Tm^2}$. Here
$\lambda_J\approx\sqrt{\Phi_0/( 4\pi\mu_0 j_c\lambda_L)}$ is the
Josephson length; $\mu_0$ is the magnetic permeability of the vacuum
and $\lambda_L$ is the London penetration depth of both electrodes.

Up to now three different types of \ZeroPi Josephson junctions
exist. One approach makes use of the $d_{x^2-y^2}$ wave order
parameter symmetry in cuprate
superconductors\cite{VanHarlingen95,Tsuei00,Chesca02,Smilde02,Ariando05}.
Another approach is to use standard $\Nb/\Al\mbox{-}\Al_2\O_3/\Nb$
Josephson junctions equipped with current
injectors\cite{Ustinov02,Goldobin04} which allow to create any phase
shift. \ZeroPi Josephson junctions were also produced
(accidentally) by SFS technology\cite{DellaRocca05,Frolov06}. The
first intentionally made \ZeroPi SIFS junction including reference $0$ and $\pi$ Josephson junctions fabricated in the same run were recently
realized\cite{Weides06a}. Some static and dynamic properties of this
type of \ZeroPi junction were
studied experimentally\cite{Weides06a,WeidesJAP07,Pfeiffer08}. Relevant theoretical work on SIFS junctions can be found in \cite{Vasenko08,Volkov09}.

The aim of the present paper is to provide a careful analysis of the
magnetic field dependence of the junction critical current $I_c(B)$ in order to characterize these novel type of junctions as accurately as possible.
The (short) junction we discuss has a length $L \approx \lambda_J$.
As we will see, the measured $I_c(B)$ can be reproduced very well
when, apart from asymmetries of the critical current densities in the $0$ and $\pi$ parts, asymmetric flux penetration into the $0$ and $\pi$ parts
is taken into account.

The paper is organized as follows: In section \ref{Sec_Exp} the SIFS junctions are characterized in terms of geometry, and the properties of the F-layer are further characterized by measuring the magnetization of a bare $\Ni_{0.6}\Cu_{0.4}$ thin films with thickness comparable to the F-layer used for the junctions. In the central section \ref{Sec_Ich} the magnetic field dependence of the critical current of the SIFS junctions is discussed. Section \ref{Sec_Conclusion} contains the conclusion.

\section{Sample characterization}\label{Sec_Exp}

\begin{figure}[tb]
\begin{center}
  \includegraphics[width=8.6cm]{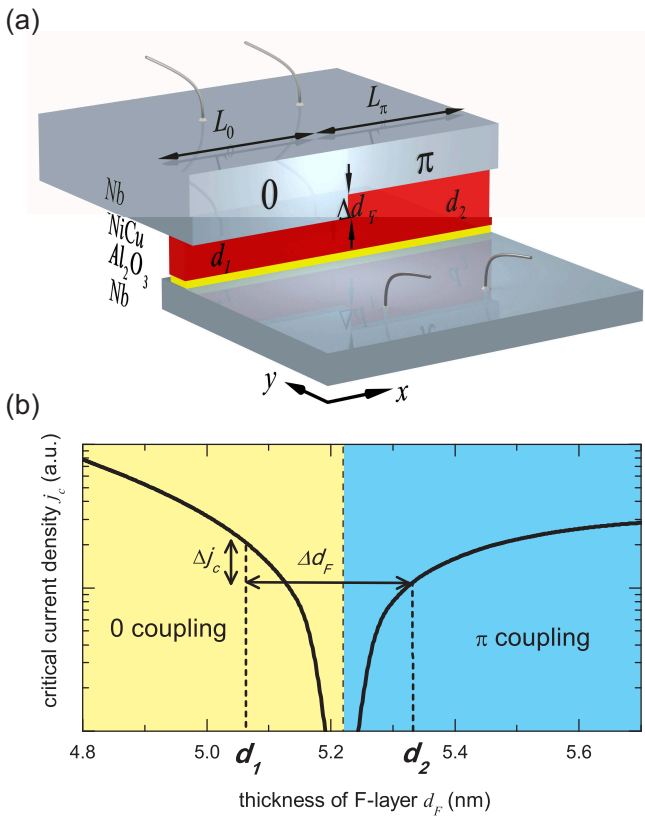}
  \caption{(Color online) (a) Sketch of a \ZeroPi SIFS Josephson junction with step-like F-layer
  to create a $0$-coupled part (F-layer thickness $d_1$) and a $\pi$ part ($d_2$). $L_0,L_{\pi}$ denote the length in each part.
  (b) Schematic $j_c(d_F)$ dependence for SIFS Josephson junctions. For the ferromagnetic thicknesses $d_1$ and $d_2=d_1+\Delta d_F$ the critical current densities $j_c(d_1)$ and $j_c(d_2)$ have similar absolute values ($j_c(d_1)=j_c(d_2)+\Delta j_c$).}
  \label{Fig:SketchopiJJ}
\end{center}
\end{figure}

Fig.~\ref{Fig:SketchopiJJ}(a) shows a sketch of the \ZeroPi
junction used in the experiment. The superconducting bottom and top
layers consist of Nb with the thicknesses $t_1=120\;\rm{nm}$ and
$t_2=400\;\rm{nm}$, respectively. As for standard Nb tunnel
junctions an $\Al_2\O_3$ layer was used as tunnel barrier. Its
thickness is $d^{J}_I\approx 0.9\;\rm{nm}$, determined from dynamic
measurements. For the ferromagnetic layer we use the diluted
ferromagnet $\Ni_{0.6}\Cu_{0.4}$. To form a \ZeroPi junction the
junction is divided into two parts differing by the thickness of the
F-layer. While in one half of the junction the thickness $d_1$ is
chosen such that $0$ coupling is realized, in the other half the F
layer thickness $d_2$ is used to realize $\pi$ coupling. In order to
have approximately symmetric junctions, $d_1$ and $d_2$ should be
such that the critical current densities of the two halves are about
the same and as large as possible, see Fig.~\ref{Fig:SketchopiJJ}(b).

Details of the fabrication technique can be found in Refs.
\onlinecite{Weides07, WeidesJAP07}. The main
feature is a gradient in the ferromagnetic $\Ni_{0.6}\Cu_{0.4}$ layer
along the $y$ direction of the 4'' wafer, in order to allow for a
variety of $0$ and $\pi$ coupled junctions differing in their critical
current densities. In addition, by optical lithography and
controlled etching, parts of the F-layer are thinned by $\Delta
d_F\approx 3\rm\AA$, such that $0$ coupling is achieved in these
parts. Thus the chip contains un-etched parts with F-layer thickness
$d_F(y)$, as well as uniformly etched parts with F-layer thickness
$d_F(y)-\Delta d_F$. Thus, at a fixed $y$-position we have two different
ferromagnetic thicknesses allowing for patterning a set of three
junctions:
\begin{enumerate}[$\bullet$]
\item a $0$ junction with F-layer thickness $d_{1}$ and critical current density \nolinebreak $j_c^0\equiv \nolinebreak  j_c(d_1)$
\item a $\pi$ junction with F-layer thickness $d_{2}$ and critical current density $j_c^\pi\equiv j_c(d_2)$
\item a stepped \ZeroPi junction with thicknesses $d_1$, $d_2$ and critical densities $j_c^0$, $j_c^\pi$ in $0$ and $\pi$
halves.
\end{enumerate}
For the values $d_1=5.05\;\rm{nm}$ and $d_2=5.33\;\rm{nm}$ we
achieved $j_c^0\approx 2.1 \; \rm A/cm^2$ and $j_c^\pi\approx 1.7 \;
\rm A/cm^2$ at $T=4.2\,\rm K$, as estimated from $0$ and $\pi$
reference junctions. Due to the different temperature dependence of
$j_c^0$ and $j_c^\pi$ (see Ref.~\onlinecite{Weides06}) these values change to
$j_c^0\approx j_c^\pi\approx 2.2 \; \rm A/cm^2$ at $T=2.65\,\rm K$.

\begin{figure}[tb]
\begin{center}
  \includegraphics[width=6.6cm]{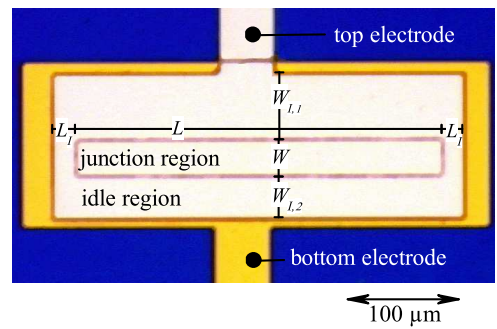}
  \caption{(Color online) Optical image (top view) of a $330\times30\;\rm{\mu m^2}$ window junction with junction and idle regions.}
  \label{Fig:ImageopiJJ}
\end{center}
\end{figure}

All junctions had the same geometrical dimensions $330 \times 30
\;\rm{\mu m^2}$, see Fig.~\ref{Fig:ImageopiJJ}. The
superconducting electrodes extend well beyond the junction area,
leading to an idle region around the junction affecting the
Josephson length $\lambda_J$. Ignoring this correction, using
$j_c^0=j_c^\pi=2.2 \; \rm A/cm^2$, as measured at $T=2.65\,\rm K$,
one finds $\lambda_J\approx 260\;\mathrm{\mu m}$, i.e. $L\approx 1.2 \lambda_J$ as in Ref.~\onlinecite{Weides06a}. The
idle region of width $W_{I,1}+W_{I,2}$ in $y$ direction leads to an
effective Josephson length\cite{Monaco1995}
\[\lambda_{J,\mathrm{eff}}=\lambda_J\sqrt{1+\frac{W_{I,1}+W_{I,2}}{W}\frac{d'_J}{d'_I}},\]
with the junction width $W$, and the inductances (per square) of the superconducting films forming the junction electrodes
$\mu_0d'_J$ and the idle regions $\mu_0d'_I$. For our junction we get
$\lambda_{J,\mathrm{eff}}=1.7 \lambda_J$, with $W=30\,\rm{\mu m}$,
$W_{I,1}+W_{I,2}=100\,\rm{\mu m}$, $d'_J =194 \,\rm{nm}$, and
$d'_I=350 \,\rm{nm}$. Therefore the normalized junction length at
$T=2.65\, \rm{K}$ is $l=L/\lambda_{J,\mathrm{eff}}\approx 0.76$ and
we clearly are in the short
junction limit.\\

\subsubsection*{Magnetic properties of the F-layer}
\begin{figure}[tb]
\begin{center}
\includegraphics[width=8.6cm]{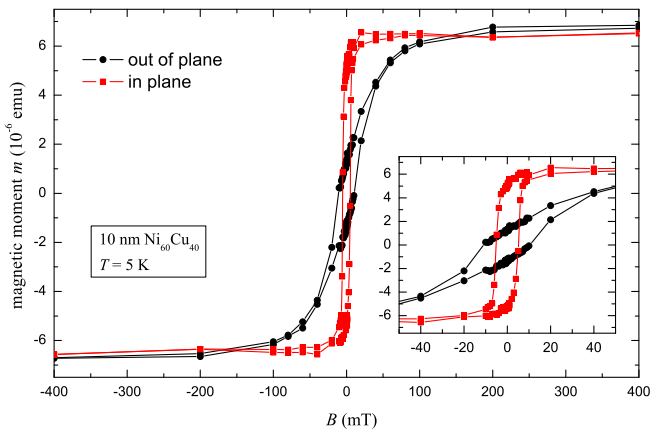}
\caption{\label{Fig:SQUID}(Color online) Magnetization curves of a $\Ni_{0.6}\Cu_{0.4}$ thin film with $10\;\rm{nm}$ thickness at $T=5\;\rm{K}$ probed by SQUID magnetometry. The magnetic field was applied either in-plane (squares) and
out-of-plane (circles). The inset shows a magnification at small magnetic fields.}
\end{center}
\end{figure}

In order to investigate the magnetic properties of the
$\Ni_{0.6}\Cu_{0.4}$ alloy used for the F-layer we performed
measurements of the magnetization via SQUID magnetometry.
The sample was a $10\;\rm{nm}$ thin $\Ni_{0.6}\Cu_{0.4}$
film deposited directly on a $\Si\O_2$ substrate.
The obtained magnetization curves (after diamagnetic correction) at $T=\;5\rm{K}$ are shown in Fig.~\ref{Fig:SQUID} for the magnetic field applied in-plane or out-of plane. The magnetic moments for the out-of plane and in-plane component saturate at almost equal $m\approx6.5\times 10^{-6}\;\rm{emu}$ corresponding to a saturation magnetization $M=130\;\rm{kA/m}$. Using the density $\rho=8.92\;\rm{g/cm^3}$ (bulk value) of the F-layer and the molar weight $60.6\; \rm{g}$ we can estimate the atomic saturation moment $m_\mathrm{at}=0.16 µ_B$, in good agreement with $m_\mathrm{at}=0.15 µ_B$ found in literature \cite{AhernNiCu}.\\
In the inset of Fig.~\ref{Fig:SQUID} the hysteresis of the magnetization curves is shown at small applied magnetic fields. Remanence can be seen for the in-plane as well as the out-of plane curves. The inversion of the magnetizations is smooth, indicating a multiple domain state. The magnetic field necessary to fully magnetize the magnetic film in-plane is in the order of $10 \;\rm{mT}$, whereas the out-of plane magnetization saturates above about $100\;\rm{mT}$. Therefore we expect the in-plane magnetization to be energetically favorable.\\
Both saturation fields are orders of magnitude larger than the in-plane fields typically used for SIFS critical current versus magnetic field measurements. In the following we estimate an upper limit by
how much the $I_c(B)$ pattern (of a $0$ junction or a $\pi$
junction) would shift along the field axis for an in-plane, fully
saturated ferromagnetic layer. Our measured saturation magnetization $M=130 \;\rm{kA/m}$ yields a magnetic induction $\mu_0 M= 0.163\;\rm{T}$. A cross section of length $L$ and a thickness $d_F$ encloses an intrinsic
magnetic flux $\Phi_M=d_F\cdot L\cdot \mu_0 M$.
For $L=330\:\rm{\mu m}$ and
 $d_F=5\:\rm{nm}$ the magnetic flux is
${\Phi_M}=129\times{\Phi_0}$. Thus, the $I_c(B)$ pattern would be shifted along the field axis
by about $129$ periods, while in experiment typically shifts of much
less than one period are observed. Further, nearly all our SIFS
junctions had mirror-symmetrical $I_c(B)$ patterns for
$|B|<1\;\rm{\rm{m T}}$, again strongly indicating that the F-layer
is in a multiple domain state with a very small in-plane net
magnetic flux \cite{Weides08}. The out-of-plane net magnetic flux has
to be small too. As we will see in the next section, for the $0$
and $\pi$ junctions highly symmetric $I_c(B)$ patterns can be
measured. If the out-of-plane magnetic flux were very large, one
would expect a large number of Abrikosov vortices penetrating the
superconducting layers, making the $I_c(B)$ of SIFS junctions with a
planar F-layer strongly asymmetric.\\
The ferromagnetic properties of a comparable ferromagnetic
compound, $\Cu_{0.47}\Ni_{0.53}$, were investigated recently via anomalous Hall voltage
measurements and Bitter decoration techniques of the magnetic domain
structures \cite{Veshchunov08}, indicating a 
magnetic anisotropy and a
magnetic structure with domains of about
$100\,\rm nm$ in size. Both Hall and Bitter decoration measurements are only sensitive to out-of-plane components of the magnetic fields, and the growth conditions of the $\Cu\Ni$ sample in Ref.~\onlinecite{Veshchunov08} may influence its magnetic properties. Nevertheless it supports our experimental findings of a very small in-plane magnetization for zero field cooled samples and a multiple domain state in the F-layer of our SIFS devices.\\

\section{critical current vs. magnetic field}\label{Sec_Ich}
In order to measure the magnetic field dependence of the critical
currents of our junctions, the samples were mounted in a glass-fiber
Helium cryostat surrounded by a triple mu-metal shield. To minimize
external noise the whole setup was placed in a high-frequency screnning
room, the current leads were low-pass filtered, and all electronics
within the screnning room was powered by batteries. The sample was
initially cooled from room-temperature down to $4.2\;\rm K$ with the
sample mounted inside the magnetic shield. To remove magnetic flux
sometimes trapped in the superconducting electrodes the sample was
thermally cycled to above the superconducting transition temperature
$T_c$. To determine $I_c$ we used a voltage criterion of
$V_\mathrm{cr}=0.5\;\rm \mu V$. The current-voltage ($IV$) characteristics and
$I_c(B)$ were measured for all three junctions at various
temperatures $T=4.2\ldots2.65\,\rm{K}$. The magnetic field $B$ was
applied along the $y$ direction see Fig.~\ref{Fig:SketchopiJJ}(a). \\
\begin{figure}[tb]
\begin{center}
\includegraphics[width=8.6cm]{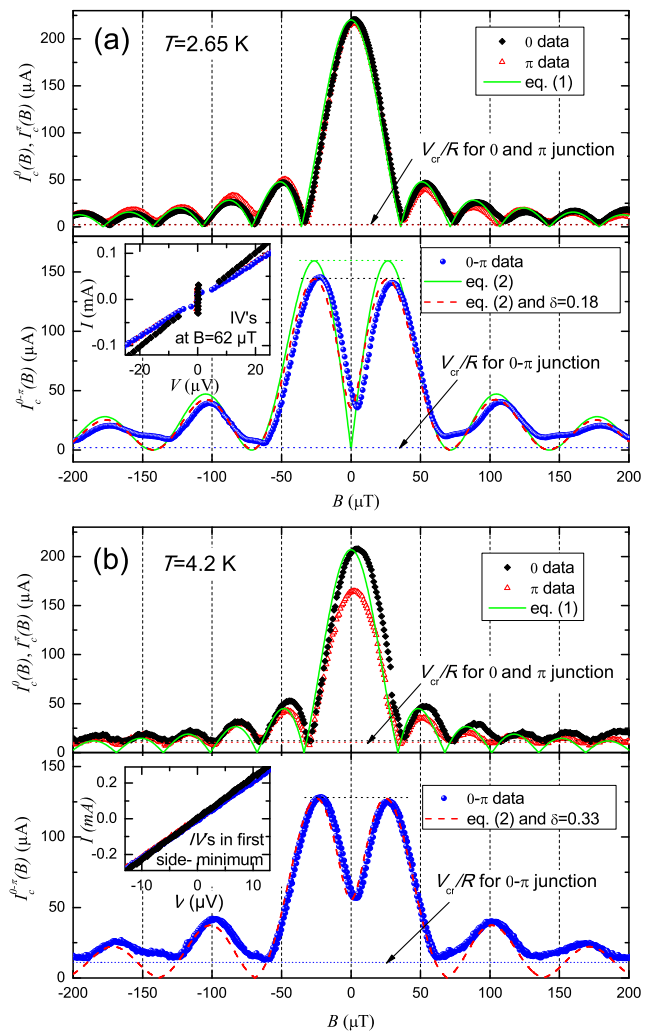}
\caption{(Color online) $I_c(B)$ measurements and theoretical curves
(short junction model) for 0, $\pi$ and \ZeroPi Josephson junctions
at (a) $T\approx2.65\;\rm{K}$ and (b) $T\approx4.2\;\rm{K}$. In the
top graphs of (a) and (b) data for the $0$ and $\pi$ junctions are
shown by solid symbols; the Fraunhofer curves Eq.~(\ref{eq_fraunhofer}) are shown by the solid lines. In the lower
graphs data for the \ZeroPi junction are shown by symbols; the solid
line corresponds to Eq.~(\ref{Eq:Fraunhofer0piJJ}). For the
theoretical curve shown by the dashed lines an asymmetry in the critical current densities
$\delta\equiv(j_c^0-j_c^\pi)/(j_c^0+j_c^\pi)=0.18$ in (a) and $0.33$
in (b) has been assumed. The horizontal dotted lines show the value of $I_\mathrm{c,min}$. The insets in (a) and (b) show $IV$-curves
for all three Josephson junctions, using the same symbols as for the
$I_c(B)$ patterns. } \label{Fig:SFIcH}
\end{center}
\end{figure}
Figure \ref{Fig:SFIcH} shows measurements of $I_c(B)$ at (a)
$T\approx2.65\;\rm{K}$ and (b) $T=4.2\,\rm{K}$. Together with the
experimental data we plot theoretical curves using the analytic
expressions valid for short junctions having homogenous critical
current density.

For the $0$ and $\pi$ junctions one has the Fraunhofer pattern:
\begin{equation}
I_c^{0,\pi}(B)=I_c^{0,\pi}(0)
\left|{\frac{\sin(\frac{\pi\Phi}{\Phi_0})}{\frac{\pi
\Phi}{\Phi_0}}}\right | \label{eq_fraunhofer}~,
\end{equation}
where $\Phi/\Phi_0=BL\Lambda/\Phi_0$ is the number of the applied
flux quanta through the normalized junction area $L\Lambda$, with $\Lambda =
d_I+d_F+\lambda_L\tanh(t_1/2\lambda_L)
  +\lambda_L\tanh(t_2/2\lambda_L)$.\\
For  a symmetric, short \ZeroPi junction the analytical expression
is given by \cite{WollmanHarlingen,Xu:SF-shape}:
\begin{equation}
  I_c^{\ZeroPi}(B) = I_c^0 \frac{\sin^2\left(\frac{\pi}{2}\frac{\Phi}{\Phi_0}\right)}{\left|\frac{\pi}{2} \frac{\Phi}{\Phi_0}\right|}
. \label{Eq:Fraunhofer0piJJ}\end{equation} \\

At $T=2.65\;\rm K$ the reference junctions have basically the same
maximum critical current of $I_c^{0}\approx 220\;\rm{\mu A}$ and
$I_c^{\pi}\approx 217\;\rm{\mu A}$ and are fitted very well by the
standard Fraunhofer curve given by Eq.~(\ref{eq_fraunhofer}). Note that the
maximum is shifted along the $B$ axis by a few percent of one flux quantum. For
reference we also show by a dotted horizontal line the
$I_c$-detection limit $I_{c,\mathrm{min}}=V_\mathrm{cr}/R$ set by the finite
voltage criterion. Here $R$ denotes the (subgap) junction resistance
at small voltage. $R$ was estimated from the corresponding
$IV$-curves shown in the insets of Fig.~\ref{Fig:SFIcH}. For the
measurements at $T=2.65\;\rm K$ this line is marginally shifted
from zero.

Looking at $I_c^{\ZeroPi}(B)$ of the stepped \ZeroPi
junction at $T=2.65\;\rm K$ (see bottom graph of Fig.~\ref{Fig:SFIcH}(a)), we see that the agreement between the
analytical expression Eq.~(\ref{Eq:Fraunhofer0piJJ}) and the
measurement is worse than for the reference junctions. For
example the central minimum of $I_c^{\ZeroPi}(B)$ is reproduced
qualitatively, however, apart from a slight shift to positive
magnetic field values, it does not reach zero critical current and
is U-shaped in contrast to the V-shaped central minimum predicted by Eq.~(\ref{Eq:Fraunhofer0piJJ}).
Further, the side maxima in $I_c^{\ZeroPi}(B)$ at the magnetic
field $\pm B_m=2\Phi_0/L\Lambda$ are below the theoretical value of
$0.72I_c^0$. Additionally we found a small asymmetry of the maxima of
$4\;\rm{\%}$, i.e. $I_c^{\ZeroPi}(-B_m)/I_c^0\approx0.66$ and
$I_c^{\ZeroPi}(+B_m)/I_c^0\approx0.64$. Finally, the first side
minima of $I_c^{\ZeroPi}(B)$ were reached at the same magnetic
field ($\Phi/\Phi_0=\pm2$) as the second minima of the $I_c(B)$ of
the reference junctions, but exhibit bumps and do not reach
zero-level defined by the $I_\mathrm{c,min}$ line.

All discrepancies to the calculated pattern, especially the non-vanishing minima, are not due
to our measurement technique. All characteristic features are well
above our $I_c$ detection limit, drawn by the dotted line in the
bottom graph of Fig.~\ref{Fig:SFIcH}(a). The U-shaped central
minimum $I_c^{\ZeroPi}(0)$ could be due to fluctuations in the applied
magnetic field. However, careful measurements using superconducting
magnetic field coils in persistent mode to exclude any magnetic field
noise showed no further decrease of the minimum. An improved fit can
be achieved by assuming that the critical current densities of the
two halves of the \ZeroPi junction are not identical, i.e. are
different from the respective $j_c^0$ and $j_c^\pi$ of the reference
junctions (e.g. caused by some gradient of the ferromagnetic
thickness along $x$ direction; the distance between reference and
stepped junctions on the chip is about $2\;\rm{mm}$). The
dashed line in the bottom graph of Fig.~\ref{Fig:SFIcH}(a) shows
the result of a corresponding calculation (the procedure is
discussed further below) using $\delta
\equiv(j_c^0-j_c^\pi)/(j_c^0+j_c^\pi)=0.18$. While the critical
current value of the central minimum is reproduced reasonably well,
the other discrepancies remain.\\
Fig.~\ref{Fig:SFIcH}(b) shows data for $T=4.2\,\rm{K}$. The
critical currents of the $0$ and $\pi$ reference junction differ by
$\approx 22\%$, but still are reasonably well described by the
Fraunhofer pattern Eq.~(\ref{Eq:Fraunhofer0piJJ}). The main
discrepancy between fit and measurements can be found in the minima
of $I_c(B)$. The experimental minima do not reach zero current,
which at this temperature is due to the finite voltage criterion,
c.f. horizontal dotted lines. The lower graph in Fig.
\ref{Fig:SFIcH}(b) shows the corresponding $I_c(B)$ measurement for
the \ZeroPi junction together with a theoretical curve, using
$\delta=0.33$. Although the overall agreement between the two curves
is reasonable, again the shape of the minima is not reproduced well.

To further discuss the observed discrepancies we either have to
assume, that $j_c^0$ and $j_c^\pi$ are non-uniform over the junction length, which would be contradictory to the observations at the reference junctions, or we should
consider effects caused by a possible remanent magnetization of the
F-layer, which can be different in the $0$ and $\pi$ part, plus the
possibility that the magnetic flux generated by the applied field
may be enhanced by the magnetic moment of the F-layer. Also, the
effective junction thickness $\Lambda$ may be different in the $0$
and $\pi$ parts, causing additional asymmetries. To account for
these effects, the local phases in the two parts may be written as
\begin{eqnarray}
\mu^0(x)=\phi_0+(\varphi^0_B+\varphi^0_M)x/L_0\\
\mu^\pi(x)=\phi_0+(\varphi^\pi_B+\varphi^\pi_M)x/L_\pi.
\end{eqnarray}
Here, $\phi_0$ is an initial phase to be fixed when calculating the
total critical current. $\varphi^{0,\pi}_M$ are the fluxes,
normalized to $\Phi_0/2\pi$, that are generated by the (1D) $y$-component of the in-plane remanent
magnetizations in the $0$ and $\pi$ parts, respectively.
$\varphi^{0,\pi}_B$ are the normalized fluxes through the junction
generated by the applied magnetic field.
In the following we parameterize $\varphi^{0,\pi}_M$ as
$\varphi^{0,\pi}_M=\bar{\varphi}_M(1\pm\delta_M)$ and $\varphi_B^{0,\pi}$ as
$\varphi_B^{0,\pi}=\bar{\varphi}_B(1\pm\delta_B)$, respectively. We further
set $L_0=L_\pi=L/2$ which is the case for the sample discussed here.

To obtain the junction critical current $I_c^{\ZeroPi}$ as a
function of the applied magnetic field, we first calculate the
currents $I_0$,$I_\pi$ in the $0$ and $\pi$ parts via
\[
I_0=\int_{-L_0}^0 j_c^0 \sin\left(\mu^0(x)\right)dx \, ,
\]
\[
I_\pi=\int_{0}^{L_\pi} j_c^\pi \sin\left(\mu^\pi(x)+\pi\right)dx\, ,
\]
and maximize $I_0+I_\pi$ with respect to $\phi_0$ for each value of
the applied magnetic field.

\begin{figure}[tb]
\begin{center}
\includegraphics[width=8.6cm]{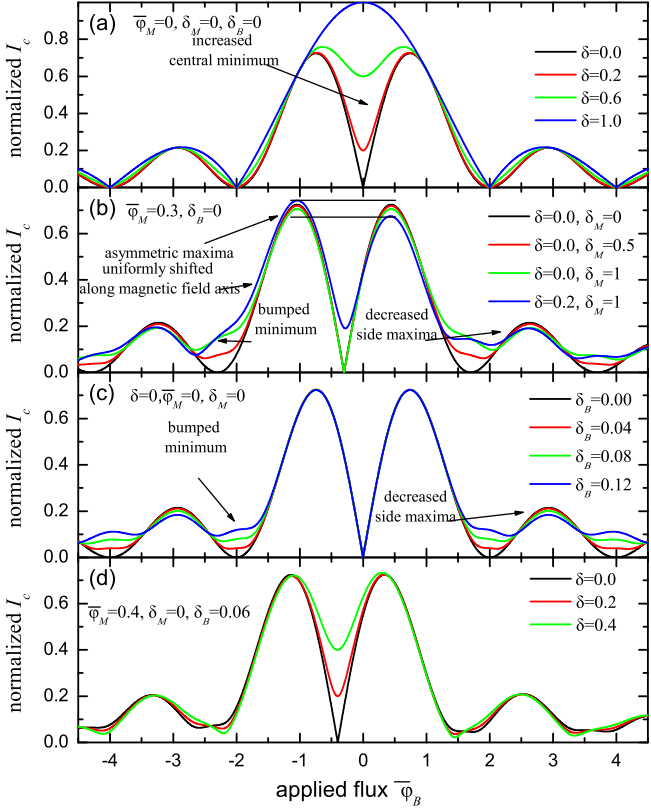}
\caption{(Color online)\label{Fig:IcHAsymLondonPenetration}
Calculated magnetic diffraction pattern
$I^{\ZeroPi}_c(\bar{\varphi_B})$ for a short \ZeroPi junction
with asymmetries in the critical current densities and in the
magnetizations of the $0$ and $\pi$ part: (a) effect of the
asymmetry parameter of the critical current density $\delta$; (b) resulting pattern with additional remanent magnetizations
(average value $\bar{\varphi}_M$ and asymmetry $\delta_M$); (c) effect of the asymmetry parameter
$\delta_B$ caused by the applied flux; (d) effect of $\delta$ for nonzero values of $\bar{\varphi}_M=0.4$ and $\delta_B=0.06$.}
\end{center}
\end{figure}
We first address the effect of the parameters $\delta$, $\bar{\varphi}_M$, $\delta_M$, and
$\delta_B$ on the $I_c^{\ZeroPi}(\bar{\varphi}_B)$ patterns, c.f. Fig.~\ref{Fig:IcHAsymLondonPenetration}(a) to (d).

If only a $j_c$ asymmetry is considered, as shown in Fig.
\ref{Fig:IcHAsymLondonPenetration}(a), using definitions $j_c^0=j_c(1+\delta)$,
$j_c^\pi=j_c(1-\delta)$, and $j_c=0.5(j_c^0+j_c^\pi)$, one finds that with
increasing asymmetry $\delta$ the central minimum increases (for
$\delta=1$ one reaches the extremum of a non-stepped junction with
length $L_0$, while the $\pi$ part becomes ``non-Josephson'' with $j_c^\pi=0$). However in all cases the first side maxima
remain symmetric and the side minima reach zero current.

Next we would like to take into account the effect of the flux
generated by remanent magnetizations. If we consider only a non-zero
magnetization, i.e. $\bar{\varphi}_M\neq0$, with all other
parameters being zero, the $I_c(\bar{\varphi}_B)$ curve gets shifted
along the field axis, since the total flux in the junction is just
the sum of applied field and magnetization. This can be seen in Fig.
\ref{Fig:IcHAsymLondonPenetration}(b) (black curve). By adding an
asymmetry $\delta_M$ the side minima get bumped and at the same time
the maxima decrease (c.f. Fig.~\ref{Fig:IcHAsymLondonPenetration}(b) red curve). However the $I_c(\bar{\varphi}_B)$ curve is still
symmetric with respect to the central minimum. This changes by
adding an additional asymmetry $\delta \neq 0$ in the critical
current densities. Now the two main maxima get asymmetric and the
side minima get bumped (blue curve).

Now we want to consider the effect of asymmetric flux in the $0$ and
$\pi$ halves, i.e. we look at $\delta_B\neq0$. In Fig.
\ref{Fig:IcHAsymLondonPenetration}(c) we show the results obtained
by increasing $\delta_B$ with the other parameters kept at zero. The
increase of $\delta_B$ leads to bumped minima and decreased
side maxima. The resulting $I_c(\bar{\varphi}_B)$ curves looks
similar to the ones shown in Fig.~\ref{Fig:IcHAsymLondonPenetration}
(b) with asymmetries in the magnetization $\delta_M$. The comparison reveals that the
$\delta_B$ parameter acts much stronger than  $\delta_M$. The
$I_c(\bar{\varphi}_B)$ curve is still symmetric with respect to the
central minimum.

In Fig.~\ref{Fig:IcHAsymLondonPenetration}(d) we add a remanent
magnetization without asymmetry, i.e. $\bar{\varphi}_M\neq0$ and
$\delta_M =0$, and allow asymmetric critical currents $\delta\neq0$. As
one can see the maxima remain symmetric whereas the minima get
slightly asymmetric.

We further note that the calculated $I_c(B)$ patterns are identical if we
simultaneously change the sign of $\delta$, $\delta_M$, and $\delta_B$. Thus
the $I_c(B)$ pattern of the \ZeroPi junction only does not allow to identify which
parameters belong to the $0$ and $\pi$ part. However the additional information
on the (temperature dependent) critical current densities of the reference junctions may allow a clear identification of $0$ and $\pi$.

Using the above findings on the parameters $\delta$,
$\bar{\varphi}_M$, $\delta_M$, and $\delta_B$ we next discuss our
experimental data. For the non-vanishing central minimum in $I_c(B)$ a
critical current asymmetry $\delta$ is required and the shift along the magnetic
field axis can solely be caused by a finite value of
$\bar{\varphi}_M$. Thus there are only two non-trivial parameters ($\delta_M$, $\delta_B$) left to reproduce the remaining features of the experimental data.

\begin{figure}[tb]
\begin{center}
\includegraphics[width=8.6cm]{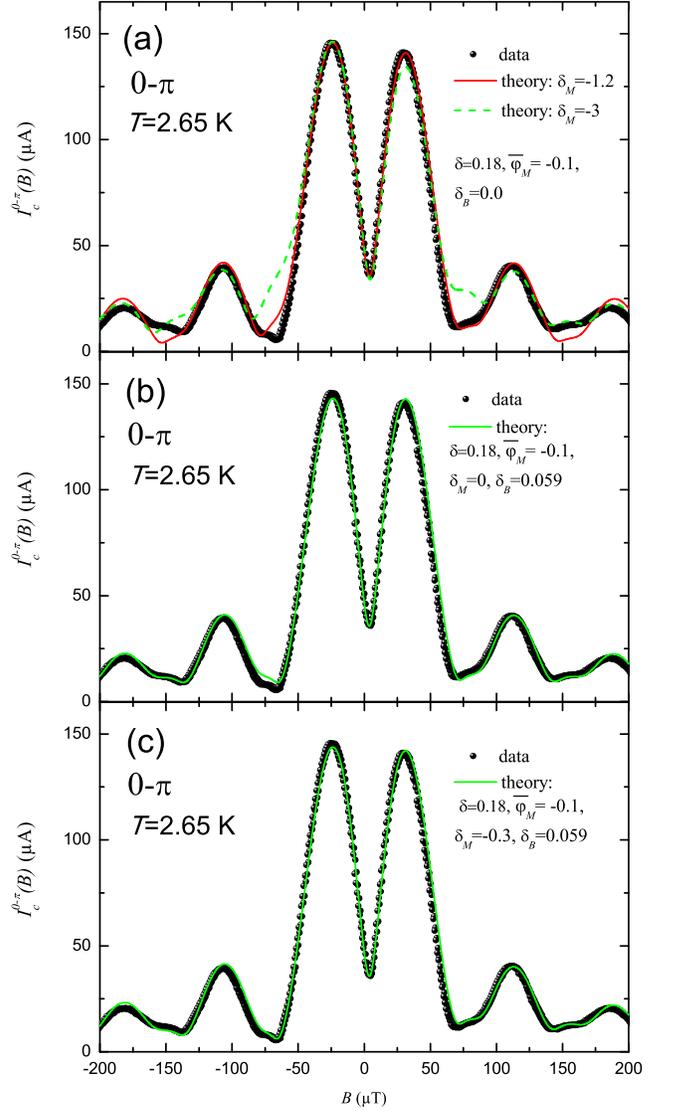}
\caption{(Color online)\label{Fig:Ic(H)_Fit} $I_c(B)$ patterns of \ZeroPi junction: comparison of experimental data and fitted magnetic
diffraction pattern $I^{\ZeroPi}_c(B)$ for $T=2.65\,\rm K$
using $\delta=0.18$ and $\bar{\varphi}_M=-0.1$. In (a) $\delta_M$ has been varied at fixed $\delta_B=0$, in (b) $\delta_B$ was
varied with fixed $\delta_M=0$, and in (c) $\delta_M$ and $\delta_B$ are varied.}
\end{center}
\end{figure}

If one allows for an asymmetry in the
remanent magnetizations only, i.e. $\delta_M\neq 0$ and $\delta_B=0$,
it is not possible to reproduce the experimental $I_c(B)$ at low and high magnetic fields at the same time.
The resulting curves can be seen in Fig.~\ref{Fig:Ic(H)_Fit}(a). For
large $\delta_M=-3$ (Fig.~\ref{Fig:Ic(H)_Fit}(a) dashed green line) the fit
works well for high fields but fails in the first
side minima. With a smaller value of $\delta_M=-1.2$ (Fig.
\ref{Fig:Ic(H)_Fit}(a) solid red line) the situation is opposite.

By contrast the parameter $\delta_B$ (with $\delta_M=0$) leads to a good agreement between the theoretical and experimental $I_c(B)$ pattern.This is shown in Fig.~\ref{Fig:Ic(H)_Fit}(b) where we used $\delta_B=0.059$. There are only small asymmetries near the side maxima and minima that cannot be reproduced for the case $\delta_M=0$. If we use both asymmetry parameters we get an excellent agreement of
the theory with the experimental data, as shown in Fig.
\ref{Fig:Ic(H)_Fit}(c) for the $T=2.65\; \rm K$ data.

\begin{figure}[tb]
\begin{center}
\includegraphics[width=8.6cm]{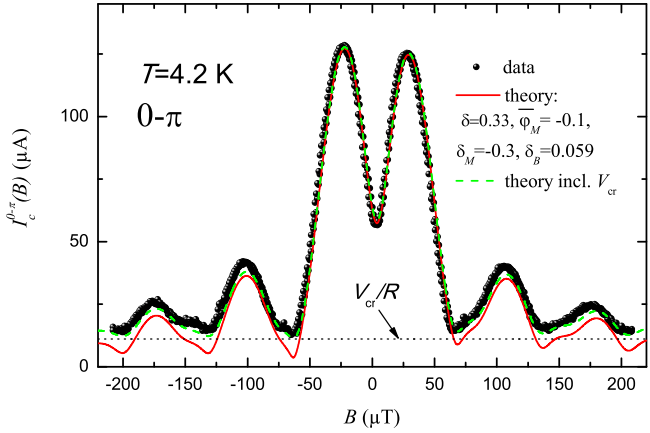}
\caption{(Color online)\label{Fig:Ic(H)_Fit42K} $I_c(B)$ patterns of \ZeroPi junction: comparison of experimental data and fitted magnetic
diffraction pattern $I^{\ZeroPi}_c(B)$ for $T=4.2\,\rm K$ (solid line). The dashed line
includes the effect of a finite voltage criterion $V_\mathrm{cr}$.}
\end{center}
\end{figure}

To further test the fit procedure we now use the
$T=4.2\;\rm K$ data and assume that the magnetic parameters remain
the same as for $T=2.65\;\rm K$. By contrast $\delta$ will change
due to a different temperature dependence of $j_c^0$ and $j_c^\pi$,
as already discussed above. For $\delta=0.33$ we get a reasonable
agreement, as shown in Fig.~\ref{Fig:Ic(H)_Fit42K}.

The $T=4.2\,\rm K$ fit is apparently not as good as the $T=2.65\,\rm K$
fit. However note that at 4.2 K the detection limit is much higher
and the minima in $I_c(B)$ are limited by the finite voltage
criterion. Still some of the bumps appear at the same values of
applied field both in the experimental and theoretical curve.

For the sake of completeness we also consider the effect of the
finite voltage criterion. Using the expression $V=R\sqrt{I^2-I_c^2}$
describing the current-voltage characteristics of a Josephson
junction in the framework of the resistively shunted junction (RSJ)
model\cite{Stewart68,McCumber68} we get a corrected $I_c(B)$ via
$I_{c,\mathrm{eff}}(B)=\sqrt{(V_\mathrm{cr}/R)^2+I_c(B)^2}$, where $I_c(B)$
refers to the theoretical curve (solid curve in
Fig.~\ref{Fig:Ic(H)_Fit42K}). The corrected curve is shown as dashed
(green) line in Fig.~\ref{Fig:Ic(H)_Fit42K}. As can be seen the data
are reproduced perfectly.

\subsubsection*{Discussion of the obtained parameters}
Finally, we would like to discuss the parameters which are obtained by fitting the experimental data.
As already mentioned above, the parameters $\delta$,
$\bar{\varphi}_M$, $\delta_M$, and $\delta_B$ allow to find the different parameter sets
for the two halves of the junction. For the distinction between ``$0$'' and ``$\pi$'' additional information is needed, which we get from the reference junctions.

The parameter $\delta$ allows to extract the absolute
values of the critical current densities in the two parts.
We get an almost temperature independent critical current density $j_c^{1}$ in
the one half with $j_c^{1}(4.2\;\mathrm{K})\approx
j_c^{1}(2.65\;\rm{K})\approx 2.3 \;\rm A/cm^2$. By contrast, for the
other part we find a temperature dependent current density $j_c^{2}$ with
$j_c^{2}(4.2\;\rm{K})\approx 1.2 \;\rm A/cm^2$ to
$j_c^{2}(2.65\;\rm{K})\approx 1.6 \;\rm A/cm^2$.
A comparison with the temperature dependencies of the reference junctions
allows the identification that the first part has to be $0$ coupled, whereas the
second part is $\pi$ coupled.
The absolute values of $j_c^{0}$ of the $0$ and \ZeroPi junction are approximately the same whereas $j_c^{\pi}$
of the \ZeroPi junction is reduced by $\approx 0.5 \;\rm A/cm^2$ as compared to the $\pi$ reference junction, although the temperature dependence looks very similar. This indicates a slightly reduced thickness of the F-layer in the $\pi$ part of the \ZeroPi junction, c.f. Figure \ref{Fig:SketchopiJJ}(b). Taking the data of Ref.~\onlinecite{Weides06} the difference in thickness can be estimated to be $\approx 0.7\;\mathrm{\AA}$. This may be caused by some gradient of the ferromagnetic thickness along $x$ direction on the chip, as the distance between reference and stepped junctions on the chip is about $2\;\rm{mm}$.

The parameters related to a different remanent magnetization in the $0$ and $\pi$ part, i.e. $\bar{\varphi}_M=-0.1$ and $\delta_M=-0.3$, seem reasonable. The magnetization is of the order of $10^{-3}$ of a fully saturated magnetization,
indicating that the F-layer is in a multi-domain state. Note that the resulting magnetization of the $\pi$ part is larger than the magnetization of
the $0$ part, which seems realistic due to a thicker F-layer in the $\pi$ part. In fact, the ratio of the F-layer thicknesses $d_2/d_1$ is very close to 1, so, assuming that magnetization is proportional to the volume of the F-layer in each part, it is quite difficult to explain the above value of $\delta_M$. However, if one assumes that there is a dead layer of thickness $d_\mathrm{dead}$ one can calculate its value from
\[\frac{d_1-d_\mathrm{dead}}{d_2-d_\mathrm{dead}}=\frac{1+\delta_M}{1-\delta_M}\]
to be $d_\mathrm{dead}\approx4.7\;\mathrm{nm}$. This value is somewhat larger than $d_\mathrm{dead}\approx3.1\;\mathrm{nm}$ estimated earlier from a $j_c(d_F)$ fit made for a different run of the same fabrication process\cite{Weides06}. However, as we see from Figs.~\ref{Fig:Ic(H)_Fit} (b) and (c) the change in $\delta_M$ from $0$ to $-0.3$ affects only the tiny features on the $I_c(B)$ curve. Thus, the value of $\delta_M$ cannot be found from this fit very exactly.

Besides the current asymmetry $\delta$ the most important parameter for our experiment is the asymmetry parameter $\delta_B$. Using a finite $\delta$ and $\delta_B=0.059$ almost all features could be reproduced very well. The addition of the parameters related to remanent magnetizations lead to minor improvements in the agreement of theory and experiment. In the following we want to discuss three possible scenarios causing the asymmetry $\delta_B$.

First, the effect could be caused simply via the fabrication procedure of the junction. In the $0$ part of the junction the SF bilayer was deposited in situ whereas the Nb cap layer in the $\pi$ part was deposited after an etching process. Thus the properties, such as the mean free path and hence the London penetration depth $\lambda$, of the Nb cap layers in the two halves could easily differ by few percents.

Second, one could think of a paramagnetic component in the magnetization. As already discussed above, the F-layer is expected to be in a multi-domain state with a small net magnetization in-plane. An external field applied in-plane could cause a reconfiguration of the domains. In the two halves the pinning of the domains may be different, due to the different thicknesses and the different treatment. This would result in a asymmetric field dependent magnetization.

A third possibility is the appearance of an enhanced flux penetration due to
inverse proximity effect, causing a correction in the London
penetration depth. Due to the reduction
of the order parameter in the vicinity of the ferromagnetic layer,
the effective penetration depth might be enlarged. In order to estimate this effect, we calculated numerically the space-dependent superfluid density $n_s(z)$ in the superconducting and the ferromagnetic part of the SF bilayer using the quasiclassical approach\cite{Vasenko08}. Herein we used the parameters of our SF bilayer, which were already obtained in Ref.~\onlinecite{Vasenko08} by fitting the experimental data of Ref.~\onlinecite{Weides06}
. By using the (London) expression $\lambda(z)\propto n_s(z)^{-0.5}$ we obtained the spatial
dependence of the penetration depth. Then we used the second London
equation $\nabla^2 B(z)=B(z)/\lambda^2(z)$ to calculate the magnetic field $B(z)$
numerically. We define the effective penetration depth as $ \lambda_{L,\mathrm{eff}}\equiv\lambda_L\Phi_{\mathrm{eff}}/\Phi$, with $\Phi_{\mathrm{eff}}$ and $\Phi$ being the flux in our SF
bilayer with and without inverse proximity corrections. For our SIFS
junctions with a thickness $d_F\approx5\;\rm{nm}$ of the ferromagnet
and $t_2=400\;\rm{nm}$ of the top electrode we get
$\lambda_{L,\mathrm{eff}}=1.005\lambda_L$ at $T=2.65\;\rm{K}$. Therefore in our case the
inverse proximity corrections are negligible. In addition the corrections due to inverse proximity effect would be opposite in sign, i.e. $\delta_B<0$, in contrast to $\delta_B=+0.059$ found for our junction.

By looking at the other two scenarios it seems natural that
the fabrication procedure causes the observed $\delta_B$ asymmetry.
However at the moment, we cannot exclude a field-dependence of the magnetization.
A clarification deserves further investigations.

\section{Conclusions}\label{Sec_Conclusion}
In this paper we presented a detailed analysis of the magnetic field
dependence in the critical current, $I_c(B)$, in $0$, $\pi$, and
\ZeroPi SIFS Josephson junctions. The length of the junctions is
smaller than the Josephson length. The $I_c(B)$ pattern of the $0$ and the $\pi$
junction can be well described by the standard Fraunhofer pattern,
valid for a homogenous, short junction. The central maximum of this
pattern is typically shifted from zero by some percent of one flux
quantum, pointing to a weak in-plane magnetization of the F-layer.
The magnetization is of order of $10^{-3}$ of a fully saturated magnetization,
indicating that the F-layer is in a multi-domain state.

The $I_c(B)$ pattern of the \ZeroPi junction exhibits the central minimum,
well known for this type of junction. However the critical current
at this minimum is non-zero, pointing to an asymmetry in the
critical current densities in the two halves of the junction. In
addition $I_c(B)$ exhibits asymmetric maxima and bumped minima that
cannot be described exclusively by critical current asymmetries. A
detailed explanation of these features requires the consideration of
asymmetric fluxes generated in the $0$ and $\pi$ parts of the junction.
A careful analysis of the experimental data and our model showed that
the majority of the observed discrepancies are due to a field-dependent
asymmetry of the fluxes in the $0$ and $\pi$ part. The effect could either be caused by a small, field-dependent, in-plane magnetization
of the F-layer or by a difference in the penetration lengths, which most naturally
can be due to the fabrication technique. In principle, this effect should also be
present in the $I_c(B)$'s of the reference junctions. However, here the effect
only leads to a small scaling factor for the magnetic field, which is too small
to be detectable in experiment, e.g. if the effects of field focusing are considered.

The model discussed in this paper on the basis of \ZeroPi junctions can be extended, e.g.,
to SIFS junctions having step-like $j_c(x)$ profile\cite{WeidesIEEE}, or laterally ordered ferromagnetic domains.

\subsection*{Acknowledgment}
This work is supported by the Deutsche Forschungsgemeinschaft
(DFG) via the SFB/TRR 21 and project GO 944/3. M. Kemmler acknowledges support by the Carl-Zeiss
Stiftung. M. Weides is supported by DFG project WE 4359/1-1.
\bibliography{PRB_bib}

\end{document}